\documentclass[12pt]{article}
\usepackage{blindtext}
\usepackage{amsmath}
\usepackage{amssymb}
\usepackage{wrapfig}
\usepackage{multicol}
\usepackage[margin=1in]{geometry}
\usepackage{graphicx}
\usepackage{caption}
\usepackage{indentfirst}

\graphicspath{{Figures/}}
\emergencystretch\maxdimen
\renewenvironment{abstract}
{\par\noindent\textbf{\abstractname:}\ \ignorespaces}
{\par\medskip}

\makeatletter
\renewcommand\maketitle
{\noindent
{\large\bfseries\@title}%
\medskip\par
{\@author}%
\medskip\par
{\@date}%
\bigskip\par\noindent
}

\newenvironment{figurehere}
{\def\@captype{figure}}
{}

\renewcommand\section{\@startsection {section}{1}{\z@}%
{-1.5ex}%
{0.05ex }%
{\bfseries}}
\makeatother

\begin{document}
\onecolumn
\title{Euclidean Quantum Field Theory from Variational Dynamics}
\author{Brenden McDearmon }
\maketitle

\begin{abstract}
A variational phase space is constructed for a system of fields on Euclidean space with periodic boundary conditions. An extended action functional is defined such that the Euler-Lagrange equations generate a symplectic flow on the variational phase space. This symplectic flow is numerically integrated as it evolves with respect to the variational parameter. Assuming ergodicity, the resulting flow samples the Euclidean path integral.

\end{abstract}

\begin{multicols*}{2}
\section*{Introduction}

Euclidean field theory is a useful tool for non-perturbative calculations of quantum systems \cite{roepstorff2012path}\cite{montvay1994quantum}. The practice of Euclidean field theory involves generating a set of field configurations on Euclidean space that sample the Euclidean path integral \cite{roepstorff2012path}\cite{montvay1994quantum}. The n-point correlation functions of the Euclidean path integral (i.e. Schwinger functions) can be used to construct relativistic n-point correlation functions (i.e. Wightman functions) on Minkowski space \cite{roepstorff2012path}\cite{montvay1994quantum}\cite{glimm2012quantum}\cite{streater2000pct}\cite{osterwalder1973axioms}\cite{osterwalder1975axioms}\cite{kondo1984constructive}\cite{frohlich1974schwinger}\cite{frohlich1977schwinger}\cite{eckmann1979time}\cite{schwinger1958euclidean}. The Wightman functions contain all of the important information of the relativistic quantum field theory, and a relativistic quantum theory can be defined by, and constructed from, an appropriate Euclidean path integral \cite{roepstorff2012path}\cite{montvay1994quantum}\cite{glimm2012quantum}\cite{streater2000pct}\cite{osterwalder1973axioms}\cite{osterwalder1975axioms}\cite{kondo1984constructive}\cite{frohlich1974schwinger}\cite{frohlich1977schwinger}\cite{eckmann1979time}\cite{schwinger1958euclidean}\cite{jost1965general}.

Sets of Euclidean field configurations can be generated on a discretized lattice using a variety of algorithms \cite{roepstorff2012path}\cite{montvay1994quantum}. Typically, Monte Carlo methods are used; however, other methods based on Langevin equations and classical dynamics are known \cite{roepstorff2012path}\cite{montvay1994quantum}\cite{damgaard1987stochastic}\cite{nelson1966derivation}\cite{batrouni1985langevin}\cite{berges2007lattice}\cite{callaway1983lattice}\cite{polonyi1983microcanonical}. One approach used in molecular dynamics simulations that appears not to have been applied yet to Euclidean field theory is a technique based on an extended phase space \cite{nose1984unified}\cite{hoover1985canonical}\cite{andersen1980molecular}\cite{ccaugin1991molecular}\cite{bond1999nose}\cite{sturgeon2000symplectic}. Extended phase space techniques introduce additional variables to the Hamiltonian to directly generate dynamics that, assuming ergodicity, sample different ensemble distributions such as the canonical ensemble, the NpT ensemble, or the  $\mu$VT ensemble \cite{nose1984unified}\cite{hoover1985canonical}\cite{andersen1980molecular}\cite{ccaugin1991molecular}\cite{bond1999nose}\cite{sturgeon2000symplectic}. In this article, an extended phase space technique is used to generate variational dynamics for a system of Euclidean fields to, assuming ergodicty, sample the Euclidean path integral. 

\section*{Variational Dynamics}

Let $ \Psi^{(m)}: x \in \mathcal{M} \mapsto \mathbb{R}$ be real components of a collection of matter and gauge fields indexed by the positive and finite integer m on d-dimensional Euclidean space $\mathcal{M}$, and let $\pi^{(\Psi)}_{m}: x \in \mathcal{M} \mapsto \mathbb{R}$ be the ``variationally conjugate'' fields. Together with a global scalar $s$ and its variational conjugate $\pi^{(s)}$, $ \gamma=\{\pi^{(\Psi)}_{m}, \Psi^{(m)}, \pi^{(s)}, s\} \in \Gamma$ is a point in the variational phase space $\Gamma$ \footnote{Since this article is directed to the computational study of discretized systems, the exact nature of $\Gamma$ in the continuum is not precisely defined. A point in the discretized $\Gamma$ is given by $\{\times_{i\in I} \pi^{(\Psi)}_{m}(i), \times_{i\in I} \Psi^{(m)}(i), \pi^{(s)}, s\}$ where i indexes the discretized fields and the index set I is finite.}. Given an action functional, $S : \gamma \in \Gamma \mapsto\mathbb{R}$, the variational dynamics is then defined by the following set of equations such that the phase space trajectory, $\gamma (\lambda) \in \Gamma$, evolves by symplectomorphisms. 
\small
 \begin{equation} \dot{\pi}^{(\Psi)}_{m}(\lambda) = \frac{d \pi^{(\Psi)}_{m}(\lambda+ \epsilon)}{d \epsilon} \vert_{\epsilon=0} = -\frac{\delta S}{\delta \Psi^{(m)}}(\lambda) \end{equation}
 \begin{equation} \dot{\Psi}^{(m)}(\lambda) = \frac{d \Psi^{(m)}(\lambda+ \epsilon)}{d \epsilon} \vert_{\epsilon=0} = \frac{\delta S}{\delta \pi^{(\Psi)}_{m}}(\lambda) \end{equation}
 \begin{equation} \dot{\pi}^{(s)}(\lambda) = \frac{d \pi^{(s)}(\lambda+ \epsilon)}{d \epsilon} \vert_{\epsilon=0} = -\frac{\partial S}{\partial s}(\lambda) \end{equation}
 \begin{equation} \dot{s}(\lambda) = \frac{d s(\lambda+ \epsilon)}{d \epsilon} \vert_{\epsilon=0} = \frac{\partial S}{\partial \pi^{(s)}}(\lambda) \end{equation}
\normalsize
While many action functionals may be defined, those studied herein are of the form $S=s \left( S^x-S^0 \right)$ where $S^x$ and $S^0$ are defined as follows.

\scriptsize
\begin{equation}
\begin{aligned} 
 S^{x}=  & \, S^v\left[\frac{\pi^{(\Psi)}_{m}}{s}\right]+\frac{(\pi^{(s)})^2}{2m_s} +\hbar\, n_f\, ln(s)+S^m[\Psi^{(m)}]
\end{aligned} 
\end{equation} 

\begin{equation}
\begin{aligned}
 S^0= S^x \vert_{\lambda=0}
\end{aligned}
\end{equation} 
\normalsize

Here, $S^v\left[\frac{\pi^{(\Psi)}_{m}}{s}\right] = \int_{\mathcal{M}} \frac{1}{2} \frac{\pi^{(\Psi)}_{m}}{s}\hat{M} \frac{\pi^{(\Psi)}_{m}}{s} dvol$ is a variational action where $\hat{M}$ is an invertable operator, and $S^m[\Psi^{(m)}]$ is a matter and gauge action. Because only discretized systems will be considered in this article, $n_f$ is the finite number of degrees of freedom for the discretized fields as defined below.

\section*{Discretization and Integration}

Consider a regular, finite periodic lattice $\mathcal{K}$. Elements of the lattice include points $p$, links $l$ connecting nearest neighbor points, plaquettes, and so on up to the dimension of the lattice. Collectively, these various lattice elements are denoted by $x \in \mathcal{K}$. The fields $\Psi^{(m)}$ and $\pi^{(\Psi)}_{m}$ are real valued n-forms which are discretized by evaluation on n-dimensional elements of the lattice, and the index m is allowed to vary with the type of lattice element $x \in \mathcal{K}$.  For example, a system consisting of a complex scalar field $\phi = \Psi^{(0)} + i \Psi^{(1)}$ is discretized by evaluation at each point $\phi(p)$ and the index m at each point p is 2 (i.e. $m\vert_{p}=2$). Similarly, a system consisting of a  u(1) gauge field $A = \Psi^{(0)}$ can be discretized by assigning a real number to each link so that the index m at each link is 1 (i.e. $m\vert_{p}=2$). The index m is allowed to vary between lattice elements of differing dimension. For example, a system consisting of a complex scalar field $\phi = \Psi^{(0)} + i \Psi^{(1)}$ and a u(1) gauge field $A = \Psi^{(2)}$ has $m\vert_{p}=2$ and $m\vert_{l}=1$. The number of degrees of freedom for the discretized fields is given by the sum $n_f=m\vert_{p} \cdot n_p + m\vert_{l} \cdot n_l + . . .$ up to the dimension of the lattice $\mathcal{K}$ where $n_p$ and $n_l$ are, respectively, the number of points and the number of links in the lattice.

Specializing to discrete variational actions of the form $S^v=\sum_x \sum_{m}  \frac{1}{2} k_x \left(\frac{\pi^{(\Psi)}_{m}(x)}{s} \right)^2$ where $k_x$ is a constant for each type of lattice element x, the discretized variational equations for the action functional $S=s \left( S^x-S^0 \right)$ defined above are given by the following where $ \mathcal{E}^{(m)}(x,\lambda)=\frac{\partial S^m}{\partial \Psi^{(m)}}(x,\lambda)$ denotes the discretized Euler-Lagrange functional.

 \begin{equation}\begin{aligned} \dot{\pi}^{(\Psi)}_{m}(x,\lambda) = - s(\lambda) \mathcal{E}^{(m)}(x,\lambda)\end{aligned}\end{equation}
 \begin{equation} \dot{\Psi}^{(m)}(x,\lambda)  =  k_x \frac{\pi^{(\Psi)}_{m}(x,\lambda)}{s(\lambda)}\end{equation}
 \begin{equation} \dot{\pi}^{(s)}(\lambda)= \left(2S^v-\hbar n_f-\left(S^x-S^0\right)\right)\vert_{\lambda} \end{equation}
 \begin{equation} \dot{s}(\lambda)=s(\lambda)\frac{\pi^{(s)}(\lambda)}{m_s} \end{equation}

Notice that the variational equations conserve the total action (i.e. $\frac{d S(\lambda+\epsilon)}{d \epsilon}\vert_{\epsilon=0} =0$ for all $\lambda$). 

Once a point in the variational phase space is specified at $\lambda=0$ (i.e. given a $\gamma(0) \in \Gamma$), the variational equations can be numerically integrated with respect to $\lambda$ to provide a trajectory in phase space $\gamma(\lambda) \in \Gamma$ using a generalized leap-frog algorithm \cite{bond1999nose}. Briefly, the numerical integration proceeds by iteration of the following assignments using a step size $\Delta \lambda$.

\scriptsize
\begin{enumerate}
\item $\pi^{(\Psi)}_{m}(x, \lambda+0.5 \Delta \lambda)=\pi^{(\Psi)}_{m}(x, \lambda)-0.5 s(\lambda) \mathcal{E}^{(m)}(x,\lambda) \Delta \lambda$ for each $x$ and m.
\item $\pi^{(s)}(\lambda+0.5 \Delta \lambda) =\frac{-2c}{\left(1+\sqrt{ (1-c \Delta \lambda / m_s)}\right)}$ \\ where $c = 0.5\Delta \lambda(n_f \hbar(1+ln(s(\lambda)))-S^v \left[\frac{\pi^{(\Psi)}_{m}(\lambda+0.5 \Delta \lambda)}{s(\lambda)} \right]+S^m\left[ \Psi^{(m)}(\lambda) \right]-S^0)-\pi^{(s)}(\lambda)$
\item $s(\lambda+\Delta \lambda)= s(\lambda)\frac{1+\frac{\Delta \lambda \pi^{(s)}(\lambda+0.5 \Delta \lambda)}{2m_s}}{1-\frac{\Delta \lambda \pi^{(s)}(\lambda+0.5 \Delta \lambda)}{2m_s}}$
\item $\Psi^{(m)}(x,\lambda+\Delta \lambda)=\Psi^{(m)}(x,\lambda)+0.5 \pi^{(\Psi)}_{m}(x,  \lambda+0.5 \Delta \lambda) \left(\frac{1}{s(\lambda+\Delta \lambda)}+\frac{1}{s(\lambda)} \right) \Delta \lambda$ for each $x$ and m.
\item $\pi^{(s)}(\lambda+\Delta \lambda)=\pi^{(s)}(\lambda+0.5\Delta \lambda)+0.5 \Delta \lambda \left(2S^v-n_f \hbar-S^x+S^0 \right)$ \\ 
where  $S^v$ and $S^x$ are evaluated with $\pi^{(\Psi)}_{m}(x,\lambda+0.5 \Delta \lambda)$, $\pi^{(s)}(\lambda+0.5 \Delta \lambda)$, $s(\lambda+\Delta \lambda)$, and $\Psi^{(m)}(x,\lambda+\Delta \lambda)$.
\item $\pi^{(\Psi)}_{m}(x, \lambda+\Delta \lambda)=\pi^{(\Psi)}_{m}(x, \lambda+0.5\Delta \lambda)-0.5 s(\lambda+\Delta \lambda) \mathcal{E}^{(m)}(x,\lambda+\Delta \lambda) \Delta \lambda$ for each $x$ and m.
\end{enumerate}
\normalsize

\section*{The Euclidean Path Integral}

Depending on the specifics of the action functional, the variational dynamics may or may not exist for all $\lambda \in [0,\infty)$. However, assuming that the variational dynamics does exist for all $\lambda \in [0,\infty)$, and further assuming that the flow is ergodic, the variational dynamics can be shown to sample the Euclidean path integral. 

By the variational equations (7) to (10), $\frac{dS}{d\lambda}=0$ so that the extended action function is conserved by the variational dynamics. Further, by definition $S|_{\lambda=0}=0$. Accordingly, the partition function for the system is defined by equation (11). 

\begin{equation}
\begin{aligned} 
\mathcal{Z} = \int \delta \left( s \left(S^x-S^0 \right)\right) d\Gamma
\end{aligned}
\end{equation}

The phase space measure $d\Gamma$ is defined as follows.
\begin{equation}
\begin{aligned} 
 d\Gamma =ds\,d\pi^{(s)}\,D\left[\pi^{(\Psi)}\right]\,D\left[\Psi\right]
\end{aligned}
\end{equation}

Here, $D[\Psi]=\prod_{m}\prod_{x} d\Psi^{(m)}(x)$ and $D\left[\pi^{(\Psi)}\right]=\prod_{m}\prod_{x} d\pi^{(\Psi)}_m(x)$. Define a change of variables $\tilde{\pi}^{(\Psi)}_{m}(x)=\frac{\pi^{(\Psi)}_{m}(x)}{s}$, and notice that $d\pi^{(\Psi)}_{m}(x) =s d \tilde{\pi}^{(\Psi)}_{m}(x)$ for s held constant. This change of variables thus results in a change of measure.
\begin{equation}
\begin{aligned} 
 d\tilde{\Gamma} =s^{n_f} ds\,d\pi^{(s)}\,D\left[\tilde{\pi}^{(\Psi)}\right]\,D\left[\Psi\right].
\end{aligned}
\end{equation}

With this change of variables, integration over the Dirac delta function with respect to $s$ can be performed using the identity $ \frac{ d }{ds} \delta\left[f(s)\right]= \frac{\delta\left[ f(s-s')\right]}{\left( \frac{df}{ds}|_{(s')}\right)}$, where $s'$ is the isolated zero of $S$ with respect to $s$. 

\scriptsize
\begin{equation}
\begin{aligned} 
s' = exp  \big[  & -\big(S^v[\tilde{\pi}^{(\Psi)}_{m}]+\frac{(\pi^{(s)})^2}{2m_s}+ S^m[\Psi^{(j)}]-S^0 \big) / ( \hbar n_f )  \big]
\end{aligned}
\end{equation}
\normalsize

Evaluating the integral with respect to s gives the following equation where $C$ is a constant.

\scriptsize
\begin{equation}
\begin{aligned} 
\mathcal{Z} =&C \int exp  \big[ -\big(S^v[\tilde{\pi}^{(\Psi)}_{m}]+\frac{(\pi^{(s)})^2}{2m_s}+ S^m[\Psi^{(m)}]-S^0 \big) / \hbar  \big] \\& d\pi^{(s)}\,D\left[\tilde{\pi}^{(\Psi)}\right]\,D\left[\Psi\right]
\end{aligned}
\end{equation}
\normalsize

 Because the integrals with respect to $d\pi^{(s)}$ and $D\left[\tilde{\pi}^{(\Psi)}\right]$ are Gaussian, they can be evaluated. Performing the integration and collecting constants provides the Euclidean path integral with constant $Z^0$.
\scriptsize
\begin{equation} 
 \mathcal{Z} = Z^0 \int exp \left [  -S^m[\Psi^{(m)}] / \hbar \right]\,D\left[\Psi\right]
\end{equation}
\normalsize

\section*{Expectation Values}

The sequence of field configurations generated by the variational dynamics can be used to calculate expectation values of observables \footnote{Here,``observables'' refer to Euclidean observables. Relativistic quantum observables may be constructed based on the Osterwalder-Schrader procedure  \cite{roepstorff2012path}\cite{montvay1994quantum}\cite{glimm2012quantum}\cite{osterwalder1973axioms}\cite{osterwalder1975axioms}.}. 

For example, for some observable $\mathcal O \left[\Psi^{(m)}\right]$ taken to be a functional of the matter and gauge fields, the expectation value with respect to the variational flow is given by the following.
\scriptsize
\begin{equation} 
\left<\mathcal O \left[\Psi^{(m)}\right]\right>_{\lambda}= \lim_{\Lambda \to \infty}\frac{1}{\Lambda} \int_0^{\Lambda} \mathcal O \left[\Psi^{(m)}(\lambda)\right] d \lambda
\end{equation}
\normalsize
The expectation value of the observable  $\mathcal O \left[\Psi^{(m)}\right]$ averaged over the variational phase space is given by the following.
\scriptsize
\begin{equation}
\begin{aligned} 
\left<\mathcal O \left[\Psi^{(m)}\right]\right>_{\Gamma}=& \frac{  \int \mathcal O \left[\Psi^{(m)}\right] exp \left [  -\left(S^m[\Psi^{(m)}] \right) / \hbar \right]\,D\left[\Psi\right]}{\int exp \left [  -\left(S^m[\Psi^{(m)}] \right) / \hbar \right]\,D\left[\Psi\right]}
\end{aligned}
\end{equation}
\normalsize

 By the assumption of ergodicity, $ \left<\mathcal O \left[\Psi^{(m)}\right]\right>_{\lambda}= \left<\mathcal O \left[\Psi^{(m)}\right]\right>_{\Gamma}$.

\section*{Examples}

In each of the following examples, a regular periodic lattice of $N=10 \times 10 \times 10 \times 10$ points $p$ having a lattice spacing of 1 was used. Complex scalar fields defined as $\phi = \Psi^{(0)}+i \Psi^{(1)}$ were assigned to the points of the lattice and were initilized with $\Psi^{(m)} = 0$ for Example 1 and  $\Psi^{(0)} = 1$ and  $\Psi^{(1)} = 0$ for Examples 2 and 4, respectively. The  fields $\pi^{(\Psi)}_0$ and $\pi^{(\Psi)}_0$ variationally conjugate to $\Psi^{(0)}$ and $\Psi^{(1)}$, respectively, were initialized with a random configuration with each $\pi^{(\Psi)}_m(p)$ randomly selected from a uniform distribution ranging from -1.75 to 1.75. The U(1) gauge field was treated at the Lie algebra level by assigning a real number to each link $l$ in the lattice. For clarity, the Lie algebra element at a link will be denoted by $A(l) = \Psi^{(m)}(l)$, and the variational conjugate field will be denoted by $\pi^{(A)}(l)$ for some $m$ depending on the example. Each $A(l)$ was initialized at 0 and each $\pi^{(A)}(l)$ was initialized by randomly selecting from a uniform distribution ranging from -1.75 to 1.75. For every example, $s$ was initialized at 1,  $\pi^{(s)}$ was initialized at 0, $\hbar$ was set equal to 1, and $m_s$ was set equal to $0.5 n_f$.

Once an initial configuration was established, each example was numerically integrated with respect to $\lambda$ using the explicit leap-frog algorithm provided by Bond, Leimkuhler, and Laird \cite{bond1999nose} and step size of $\Delta \lambda=0.01$. Each system was ``equilibrated" by stepping the system forward for 250,000 steps. After this ``equilibration,'' the system was further numerically integrated with respect to $\lambda$ for an additional 1,000,000 steps, during which data was collected.

The results provided in the examples are not intended to be quantitatively accurate. Rather, the results demonstrate the proof-of-concept use of variational dynamics for Euclidean quantum field theory calculations. Improved quantitative results could be provided by, e.g., increasing the number of points in the lattice, decreasing the step size $\Delta \lambda$, increasing the total number of steps, and further optimizing the ``equilibration'' procedure.

\subsection*{\small \;\;1.\;Complex Scalar Field With Quadratic Potential}

In this example, the complex free scalar field was simulated using a lattice discretization of $S^m= \int_M \frac{1}{2}\partial_{\mu}\phi \partial^{\mu}\bar{\phi}+ \frac{1}{2}\phi \bar{\phi}\; d^4x$ given by the following. 

\scriptsize
\begin{equation} 
\begin{aligned}
S^{m}= \sum_{p=0}^{N-1} \Bigg(\sum_{j=0}^1 & \Big(  \sum_{k=0}^3   \frac{1}{2} \left(\Psi^{(m)}(p+\mu_k)-\Psi^{(m)}(p)\right)^2 \\ &+  \frac{1}{2} \left(\Psi^{(m)}(p)\right)^2 \Big)\Bigg)
\end{aligned}
\end{equation} 
\normalsize

Here, $\mu_k$ with k = 0, 1, 2, or 3 are the four unit lattice vectors. Overall, the system is described by the following extended action functional with $n_f=2N$. 

\scriptsize
\begin{equation} 
\begin{aligned}
S=  s \Biggl[&\frac{1}{2s^2} \sum_{p=0}^{N-1} \left(\left(\pi^{(\Psi)}_0(p)\right)^2+\left(\pi^{(\Psi)}_1(p)\right)^2\right)\\ &+\frac{\left(\pi^{(s)}\right)^2}{2m_s}+n_f\hbar\, ln(s) +S^m  -S^0 \Biggr]
\end{aligned}
\end{equation} 
\normalsize

The variational dynamics was then numerically integrated with respect to $\lambda$, and the resulting development of the system was stable over the total number of steps observed.

 Figure 1 shows the development of the variational action $S^v\left[\frac{\pi^{(\Psi)}_m}{s}\right]$ (A), the matter action $S^m\left[\Psi^{(m)}\right]$ (B), the global real-valued scalar $s$ (C), and the error term $\frac{S^x-S^0}{S^0} \cdot 100\%$.

Field correlations between nearby points were calculated as $\frac{\Re\left[\left<\phi(0)\bar\phi(0+n\mu_0)\right>_{\lambda}\right]}{\left<\phi(0)\phi(0)\right>_{\lambda}}$ where $\Re$ denotes the real part and $\mu_0$ is the unit lattice vector in the 0 direction. As depicted in figure 2A, there is a small short range correlation which rapidly decays to near zero. 
\end{multicols*}
\begin{figure*}[h]
\centering
\includegraphics[width=0.95\textwidth]{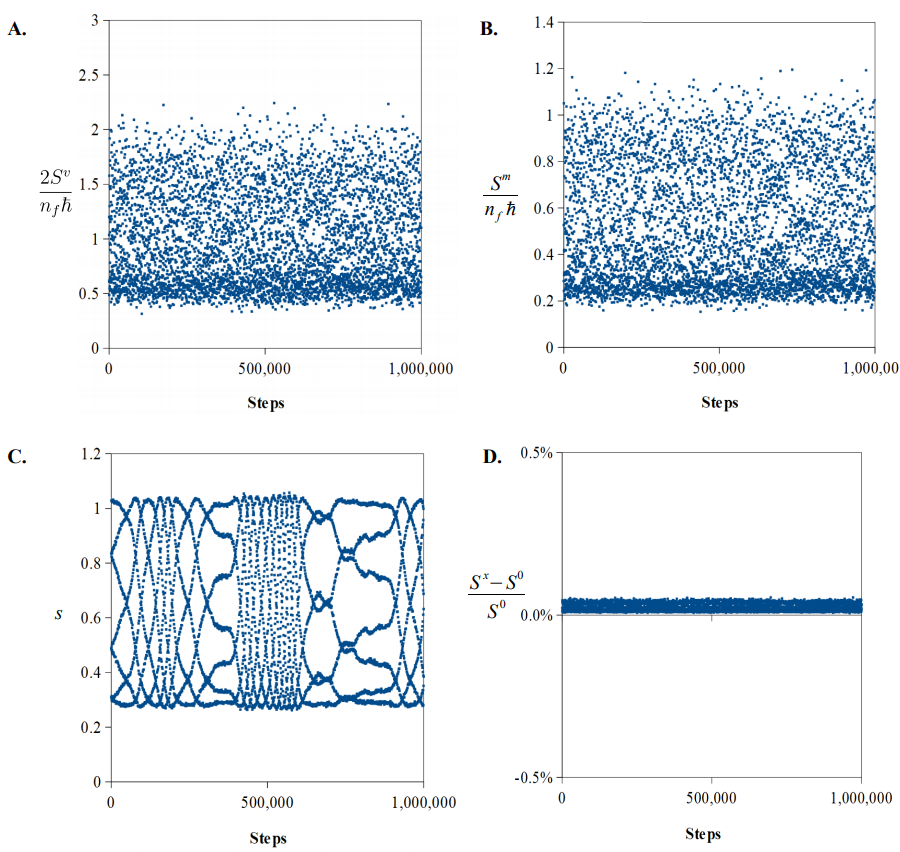}
\caption{Numerical results for Example 1 with $S^v$ (A), $S^m$ (B), s (C), and the percent error for the system (D) as a function of integration steps along $\lambda$. }
\end{figure*}
\begin{multicols*}{2}

\subsection*{\small \;\;2.\; Complex Scalar Field With Quartic Potential}
A complex  scalar field was simulated similarly as in Example 1 except a lattice discretization of a quartic matter action given by $S^m= \int_M \frac{1}{2}\partial_{\mu}\phi \partial^{\mu}\bar{\phi}- \frac{1}{2}\phi \bar{\phi}+\frac{1}{4} \left(\phi \bar{\phi}\right)^2\; d^4x$ was used. The resulting system development was stable over the total number of steps observed as seen in Figure A1 of the appendix. Comparing Figures 2A and 2B, the system with the quartic potential appears to have longer range correlations than the system with the quadratic potential. It was not determined whether these observed differences are statistically significant.

\subsection*{\small \;\;3.\; U(1) Gauge Field}

 U(1) gauge fields were treated at the Lie algebra level by assigning a real number $A(l)$ to each link $l$ in the lattice. 

\end{multicols*}
\begin{figure*}[h]
\centering
\includegraphics[width=0.95\textwidth]{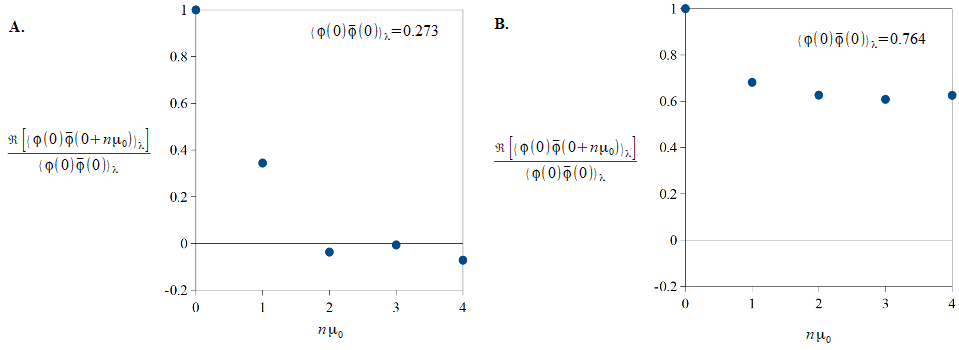}
\caption{Field correlations between nearby points for a system having a complex scalar field $\phi$ with a quadratic potential (A) and quartic potential (B).}
\end{figure*}
\begin{multicols*}{2}

The field strength term $F_{jk}(p)$ assigned to each plaquette based at point $p$ was calculated as $F_{jk}(p) =A(l_{\{p,p+\mu_j\}})+A(l_{\{p+\mu_j,p+\mu_j+\nu_k\}})-A(l_{\{p+\mu_j+\nu_k,p+\nu_k\}}-A(l_{\{p+\nu_k,p\}})$ where $\mu_j$ and $\nu_k$ are unit lattice vectors in the j and k directions respectively. The matter action was calculated as $S^m= \sum_{p=0}^{N-1} \sum_{k=j+1}^{3} \sum_{j=0}^{3} \frac{1}{2} \left(F_{jk}(p)\right)^2$. Since the matter action has a gauge degree of freedom where $S^m\left[ A+d\alpha\right] = S^m\left[ A\right]$ for any real valued scalar field $\alpha$, a gauge fixing term given by $S^{gf}$ was included. Here, the gauge fixing term $S^{gf}= \sum_{p=0}^{N-1} \frac{1}{2} \left(d^{\ast}A\right)^2(p)= \sum_{p=0}^{N-1} \sum_{j=0}^{3} \frac{1}{2} \left( A(l_{\{p,p+\mu_j\}})-A(l_{\{p,p-\mu_j\}})\right)^2$, where $d^{\ast}$ denotes the lattice exterior derivative, was used. A similar non-compact U(1) gauge theory formulation has been shown by Balaban, Imbrie, and Jaffe to be equivalent to the more traditional compact formulation where an element of the Lie group is assigned to each link \cite{balaban1984mass}.

This system was described by the extended action functional $S=s \left( S^x-S^0 \right)$ where

\scriptsize
\begin{equation} 
\begin{aligned}
&S^{x}=\\&\frac{1}{2s^2} \sum_{l=0}^{4N-1} \pi^{(A)}(l)^2+\frac{\left(\pi^{(s)}\right)^2}{2m_s}+n_f\hbar\, ln(s) +S^m[A]+S^{gf}[A].
\end{aligned}
\end{equation} 
\normalsize
Here, $n_f=4N$ and the matter action $S^m[A]$ and the gauge fixing term $S^{gf}[A]$ are described above. 

\begin{figurehere}
\centering
\includegraphics[width=0.95\linewidth]{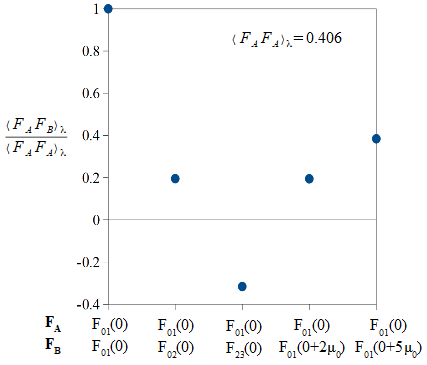}
\caption{Field strength correlations between nearby plaquettes for a system with a U(1) gauge field.}
\end{figurehere}

The variational dynamics of this system was simulated, and the resulting system development was stable over the total number of steps observed as seen in Figure A2 of the appendix. Correlations between the field strength at different plaquettes is depicted in Figure 3. Interestingly, plaquettes sharing a common link had a slight positive correlation in field strength whereas plaquettes sharing only a single common point had a slight negative correlation.

\subsection*{\small \;\;4.\; Complex Scalar Field With Quartic Potential Coupled To U(1) Gauge Field}

A system having a complex scalar field with a quartic potential coupled to a U(1) gauge field was simulated. This system was described by the extended action functional $S=s \left( S^x-S^0 \right)$ where

\scriptsize
\begin{equation} 
\begin{aligned}
&S^{x}=\\& S^v\left[\frac{\pi^{(\Psi)}_m}{s},\frac{\pi^{(A)}}{s}\right]+\frac{\left(\pi^{(s)}\right)^2}{2m_s}+\hbar\, n_f\, ln(s)+S^m[\phi, A]+S^{gf}[A].
\end{aligned}
\end{equation} 
\normalsize

Here, the variational action $S^v$ was calculated as follows.
\scriptsize
\begin{equation} 
\begin{aligned}
&S^v=\\&\frac{1}{2s^2} \left( \sum_{p=0}^{N-1} \left(\left(\pi^{(\Psi)}_0(p)\right)^2+\left(\pi^{(\Psi)}_1(p)\right)^2\right)+ \sum_{l=0}^{4N-1} \left(\pi^{(A)}(l)\right)^2 \right)
\end{aligned}
\end{equation} 
\normalsize

The matter action $S^m$ was the lattice discretization of  $\int_M \frac{1}{2}|D_A\phi|^2- \frac{1}{2}\phi \bar{\phi}+ \frac{1}{4}\left(\phi \bar{\phi}\right)^2 + \frac{1}{2}F^{jk} F_{jk} \; d^4x$ where $D_A$ is the gauge covariant derivative. The gauge fixing term $S^{gf}$ was the lattice discretization of $\int_M \frac{1}{2} \left(d^{\ast}A\right)^2 \; d^4x$ described in Example 3. The $n_f$ term is equal to 6N.

The variational dynamics of this system was simulated as described above. The resulting system development was stable over the total number of steps observed as seen in Figure A3 of the appendix. Comparing Figures 2B and 4, the local field correlations appear to decrease more rapidly with distance in the system with the gauge field (Figure 4) than in the system without the  gauge field (Figure 2B).

\begin{figurehere}
\centering
\includegraphics[width=0.95\columnwidth]{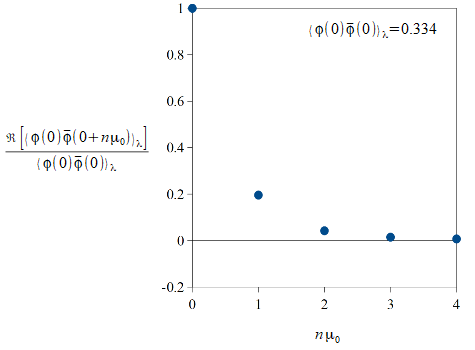}
\caption{Field correlations between nearby points for a system having a U(1) gauge field coupled to a complex scalar field $\phi$ with a quartic potential.}
\end{figurehere}

\begin{figurehere}
\centering
\includegraphics[width=0.95\columnwidth]{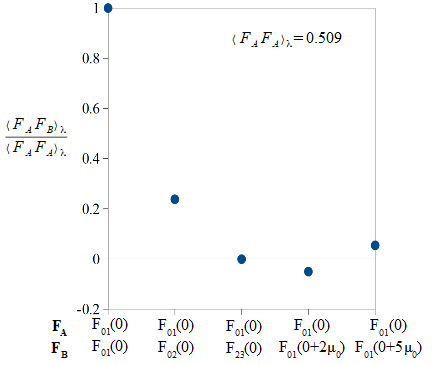}
\caption{Field strength correlations between nearby plaquettes for a system having a U(1) gauge field coupled to a complex scalar field with a quartic potential.}
\end{figurehere}

Similarly, local correlations in field strength $F_{jk}$ appear to be suppressed in the system with the gauge field couple to the complex scalar field (Figure 5) as compared to the system with the gauge field alone (Figure 3).

\section*{Discussion}

These results demonstrate the proof-of-concept use of variational dynamics for generating field configurations which, assuming ergodicity, sample the Euclidean path integral. As such, this provides a new method for performing Euclidean quantum field theory calculations. Verification of the accuracy, and assumed ergodicity, of this new technique should be conducted by, e.g., comparing these results to field configurations generated using traditional Monte Carlo methods. 

One advantage variational dynamics may have over Monte Carlo methods is the availability of new analytical techniques which make use of correlations with respect to the variational parameter. For example, one could consider various correlation functions with respect to $\lambda$ given by $C_{A,B}(\tilde{\lambda}) = lim_{\Lambda \to \infty}\int_0^{\Lambda}\mathcal O_A \left[\Psi^{(m)}(\lambda-\tilde{\lambda})\right] \mathcal O_B \left[\Psi^{(m)}(\lambda)\right] d\lambda$. In statistical mechanics, analogous correlation functions, and their Fourier transforms, are used extensively to relate dynamical properties of a system to equilibrium properties \cite{zwanzig1965time}\cite{kubo1966fluctuation}. It would be interesting to similarly apply these fluctuation-dissipation theorems to systems undergoing variational dynamics. 

Because Euclidean path integrals can be used to construct relativistic n-point correlation functions, variational dynamics may prove useful in further studying relativistic quantum field theory. For example, it may be interesting in future studies to determine whether $\lambda$-dependent aspects of variational dynamics can be preserved under the Osterwalder-Schrader mapping so that a form of variational dynamics could be constructed directly on Minkowski space or an appropriate quantum Hilbert space. Such developments may be useful in determining whether variational dynamics offers merely a new and useful tool for calculating Euclidean field configurations or whether this approach can provide further insight into the nature of quantum theory itself.

\section*{Author Contact Information}
\noindent Brenden.McDearmon@gmail.com
\bibliographystyle{unsrt}
\bibliography{citations2}
\end{multicols*}

\pagebreak
\pagenumbering{arabic}
\renewcommand*{\thepage}{A\arabic{page}}

\title{Appendix to Euclidean Quantum Field Theory from Variational Dynamics}
\author{Brenden McDearmon }
\maketitle
\\
\begin{figure*}[h]
\centering
\includegraphics[width=0.95\textwidth]{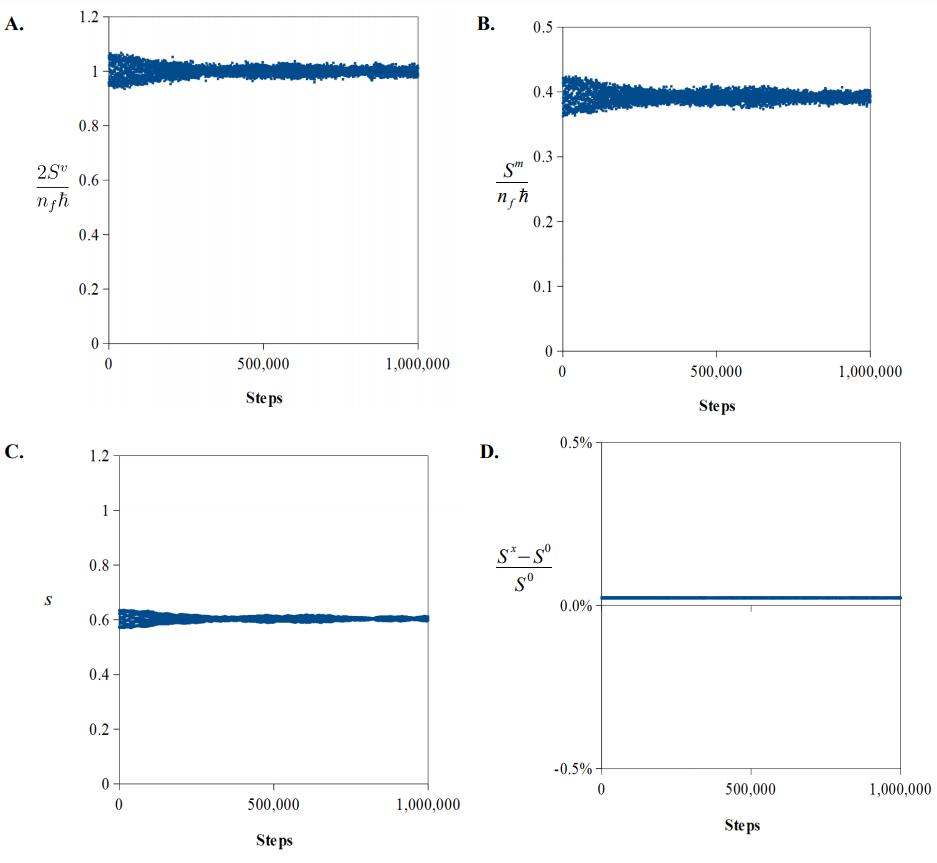}
\caption*{Figure A1: Numerical results with $S^v$ (A), $S^m$ (B), s (C), and the percent error for the system (D) as a function of integration steps along $\lambda$ for Example 2 having a complex scalar field $\phi$ with a quartic potential}
\end{figure*}

\begin{figure*}[h]
\centering
\includegraphics[width=0.95\textwidth]{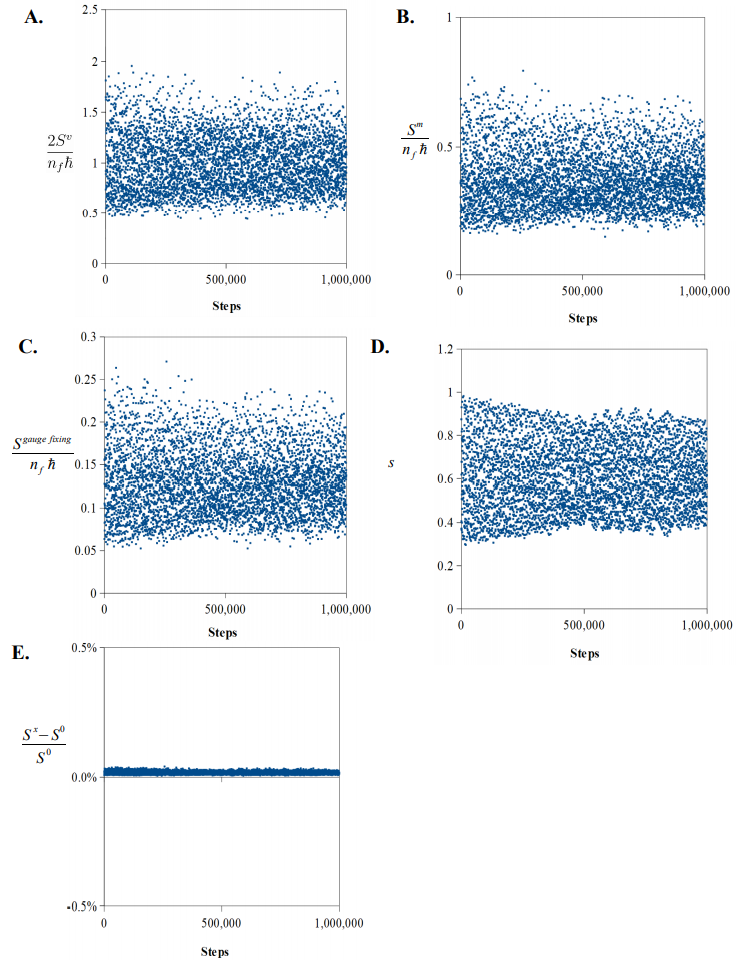}
\caption*{Figure A2: Numerical results with $S^v$ (A), $S^m$ (B), $S^{gf}$ (C), s (D), and the percent error for the system (E) as a function of integration steps along $\lambda$ for Example 3 having a U(1) gauge field with gauge fixing term $S^{gf}= \sum_{p=0}^{N-1} \frac{1}{2} \left(d^{\ast}A\right)^2(p)$.}
\end{figure*}

\begin{figure*}[h]
\centering
\includegraphics[width=0.95\textwidth]{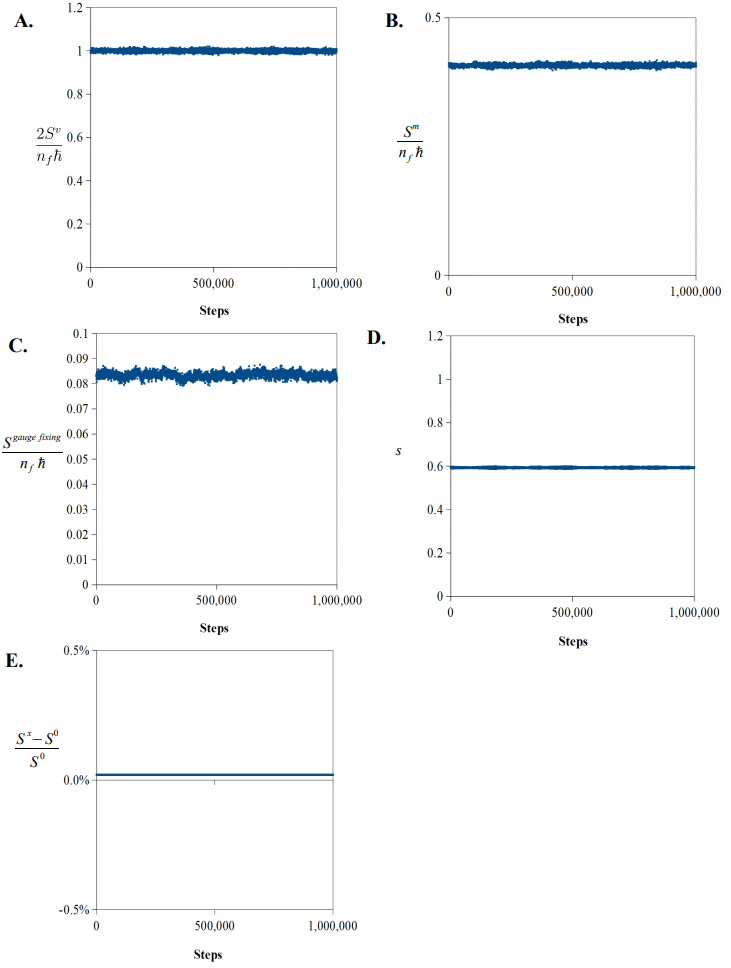}
\caption*{Figure A3: Numerical results with $S^v$ (A), $S^m$ (B), $S^{gf}$ (C), s (D), and the percent error for the system (E) as a function of integration steps along $\lambda$ for Example 4 having a U(1) gauge field with gauge fixing term $S^{gf}= \sum_{p=0}^{N-1} \frac{1}{2} \left(d^{\ast}A\right)^2(p)$ coupled to a complex scalar field $\phi$ with a quartic potential.}
\end{figure*}

\end{document}